\documentclass[letterpaper, 10 pt, journal, twoside]{IEEEtran}
\usepackage{amsmath,amssymb,euscript,psfrag,latexsym,graphicx}
\usepackage{bbm,color,amstext,wasysym,subfig,cuted,mathtools}
\usepackage[normalem]{ulem}
\graphicspath{{./},{./figures/}}

\newcommand{\mR}{{\mathbb R}}

\newcommand{\bF}{{\mathbf F}}
\newcommand{\bK}{{\mathbf K}}
\newcommand{\bP}{{\mathbf P}}

\newcommand{\bx}{{\mathbf x}}

\newcommand{\bPsi}{{\boldsymbol \Psi}}
\newcommand{\bC}{{\boldsymbol{\mathcal C}}}

\newcommand{\be}{{\mathbf e}}
\newcommand{\bX}{{\mathbf X}}
\newcommand{\bY}{{\mathbf Y}}
\newcommand{\bG}{{\mathbf G}}
\newcommand{\bA}{{\mathbf A}}
\newcommand{\bB}{{\mathbf B}}
\newcommand{\bu}{{\mathbf u}}

\newcommand{\bL}{{\mathbf L}}
\newcommand{\bd}{{\mathbf d}}
\newcommand{\bz}{{\mathbf z}}
\newcommand{\bomega}{{\boldsymbol \omega}}
\newcommand{\bpsi}{{\boldsymbol \psi}}
\newcommand{\bJ}{{\mathbf J}}
\newcommand{\bS}{{\mathbf S}}
\newcommand{\bH}{{\mathbf H}}
\newcommand{\bc}{{\mathbf c}}
\usepackage{cite}

\newtheorem{theorem}{Theorem}
\newtheorem{definition}{Definition}

\IEEEoverridecommandlockouts

\begin{document}

\title{\bf A convex data-driven approach for nonlinear control synthesis}

\author{Hyungjin Choi, Umesh Vaidya, and Yongxin Chen
\thanks{
H. Choi gratefully acknowledges funding from the Department of Energy, Office of Electricity’s Energy Storage Program, under the direction of Dr. Imre Gyuk. Sandia National Laboratories is a multi-mission laboratory managed and operated by National Technology and Engineering Solutions of Sandia, LLC., a wholly owned subsidiary of Honeywell International, Inc., for the U.S. Department of Energy’s National Nuclear Security Administration under contract DE-NA0003525. This paper describes objective technical results and analysis. Any subjective views or opinions that might be expressed in the paper do not necessarily represent the views of the U.S. Department of Energy or the United States Government.

This work is also partially supported by NSF under grant 1932458, 1901599 and 1942523, and DOE DE-OE-0000876. }
\thanks{H. Choi is with Sandia National Laboratories, Albuquerque NM, USA, hchoi@sandia.gov}
\thanks{U. Vaidya is with the Department of Mechanical Enginerring, Clemson University, Clemson SC, USA, uvaidya@clemson.edu}
\thanks{Y. Chen is with the School of Aerospace Engineering, Georgia Institute of Technology, Atlanta, GA, USA.
yongchen@gatech.edu}}

\maketitle

\begin{abstract}
We consider a class of nonlinear control synthesis problems where the underlying mathematical models are not explicitly known. We propose a data-driven approach to stabilize the systems when only sample trajectories of the dynamics are accessible. Our method is founded on the density function based almost everywhere stability certificate that is dual to the Lyapunov function for dynamic systems. Unlike Lyapunov based methods, density functions lead to a convex formulation for a joint search of the control strategy and the stability certificate. This type of convex problem can be solved efficiently by invoking the machinery of the sum of squares (SOS). For the data-driven part, we exploit the fact that the duality results in the stability theory of the dynamical system can be understood using linear Perron-Frobenius and Koopman operators. This connection allows us to use data-driven methods developed to approximate these operators combined with the SOS techniques for the convex formulation of control synthesis.
The efficacy of the proposed approach is demonstrated through several examples.
\end{abstract}

\section{Introduction}

The celebrated Lyapunov theory lays the foundation for stability analysis of nonlinear dynamical systems. Lyapunov functions provide stability certificates for a nonlinear system. For a given system, searching for a proper Lyapunov function can often be formulated as a convex optimization problem and thus easy to address. For instance, for polynomial dynamics, this is achieved through the sum of squares (SOS). Regardless of its similarity to stability analysis, the problem of nonlinear controller synthesis is more challenging. Other than a few special cases such as linear quadratic control problems, the joint search for Lyapunov stability certificate and control strategy can no longer be cast as convex optimization problems. This is exacerbated by the fact that in many applications, the underlying mathematical models are not available. Our objective in this paper is to establish a principled approach for nonlinear control synthesis when the mathematical models of the underlying dynamics are not explicitly given. 

We provide a systematic approach for data-driven control synthesis for a class of control affine nonlinear systems of the form
	\begin{equation}\label{eq:controlaffine}
		\dot \bx = \bF(\bx) + \bG(\bx) \bu.
	\end{equation}
The objective is to design state feedback controller $\bu=\bu(\bx)$ such that the closed-loop system is asymptotically stable. To achieve this objective, we use density function-based dual stability formulation introduced by Rantzer for {\it almost everywhere} stability analysis and synthesis for nonlinear control systems \cite{Ran01}. Unlike Lyapunov function-based approach for control design, the co-design problem of simultaneously finding the density function and almost everywhere stabilizing controller is a convex optimization problem. We exploit this convexity property for data-driven control synthesis. In \cite{vaidya2008lyapunov,rajaram2010stability}, it was shown that the duality between density and Lyapunov function in the stability theory could be understood using linear operator theoretic framework. In particular, the duality between Koopman and Perron-Frobenius operators is at the heart of the duality in the stability theory. This linear operator theoretic framework is also exploited for the data-driven control design \cite{das2018data,raghunathan2013optimal}.

The recent advances in the data-driven approximation of the Koopman operator are used to discover a data-driven approach for the nonlinear control synthesis. In Koopman theory, a nonlinear system is lifted to, albeit infinite-dimensional, a linear system. This lifting can be approximated using data generated from the underlying nonlinear dynamics by the well-known Extended Dynamic Mode Decomposition (EDMD) algorithm \cite{WilKevRow15}. These tools have been successfully applied in many domains, such as fluid dynamics~\cite{Mezic_2013}, power systems~\cite{Susuki_2016,Vaidya_2019}, to understand the principle components/modes of given nonlinear dynamics \cite{mauroy2016global}. Recently, Koopman theory has been introduced to the control synthesis tasks, hoping that the controller designed in the lifted space could be easier than that in the original state space. It turns out to be a challenging problem since the lifting argument in the presence of control is no longer valid. Regardless of the progress that has been made in this direction during the last few years~\cite{korda2018linear,huang2020data,kaiser2017data,kaiser2019datadriven}, a principle data-driven approach for nonlinear control synthesis is not yet available. We use the EDMD algorithm combined with the duality results for the data-driven approximation of the Perron-Frobenius (P-F) operator corresponding to the control system. This linear P-F operator for the control system is used to formulate a convex optimization problem for control synthesis. This optimization is over polynomials and can be solved using the SOS solvers. The complexity of the resulting optimization problem depends on the polynomial basis used to approximate the linear operators. Since control often doesn't require high fidelity models, we expect to construct a reliable controller using a relatively small number of basis functions. We envision that this method can be applied to low dimensional and medium dimensional dynamical systems (e.g. robotics, \textit{distributed} power-electronics control applications).

The rest of the paper is organized as follows. In Section \ref{sec:background}, we provide a review on density function methods, SOS, and Koopman theory; these are the ingredients of our approach. Problem formulation and the details of our method are presented in Section \ref{sec:main}. This is followed by several numerical examples in Section \ref{sec:case study} and a short concluding remark in Section~\ref{sec:conclusion}.

\section{Background}\label{sec:background}
Our proposed method for control synthesis utilizes density function method for control design, SOS for polynomial optimization and Koopman theory for data-driven approximations. Necessary background on these components is discussed in this section.

\subsection{Density function approach for control synthesis}
Consider control-affine system \eqref{eq:controlaffine} with feedback control $\bu(\bx)$ and $\bx \in \mathbb{R}^n$. This closed-loop system is asymptotically stable with respect to the origin $\bx=0$ if there exists a Lyapunov function $V$ such that 
	\begin{equation}\label{eq:Lyapunov}
		\frac{\partial V}{\partial \bx}^\top (\bF(\bx)+\bG(\bx)\bu(\bx)) < 0,~~\forall \bx \neq 0.
	\end{equation}
Thus, for the purpose of control synthesis, one seeks a pair $(V, \bu)$ such that \eqref{eq:Lyapunov} holds. Note that this inequality is bilinear with respect to $V, \bu$ and is thus a non-convex problem. This is the major obstacle preventing Lyapunov theory being widely used in control synthesis. In \cite{Ran01}, a dual to Lyapunov's stability theorem was established. 
\begin{theorem}[\cite{Ran01}]\label{thm:dual}
Given the system $\dot \bx = \bF(\bx)$, where $\bF$ is continuous differentiable and $\bF(0)=0$, suppose there exists a nonnegative $\rho$ is continuous differentiable for $\bx\neq 0$ such that $\rho(\bx) \bF(\bx)/|\bx|$ is integrable on $\{\bx\in \mR^n\,:\, |\bx|\ge 1\}$ and
	\begin{equation}\label{eq:dualLyap}
		[\nabla \cdot (\rho \bF)](\bx)>0~\mbox{for almost all}~\bx.
	\end{equation}
Then, for almost all initial states $\bx(0)$, the trajectory $\bx(t)$ tends to zero as $t\rightarrow \infty$. Moreover, if the equilibrium $\bx=0$ is stable, then the conclusion remains valid even if $\rho$ takes negative values.
\end{theorem}

The density $\rho$ serves as a stability certificate and can be viewed as a dual to the Lyapunov function \cite{Ran01}. Applying Theorem \ref{thm:dual} to the closed-loop system we arrive at
	\begin{equation}\label{eq:dualLyapcontrol}
		\nabla \cdot (\rho (\bF+ \bG \bu)) >0~\mbox{for almost all}~\bx.
	\end{equation}
The control synthesis becomes searching for a pair $(\rho, \bu)$ of functions such that \eqref{eq:dualLyapcontrol} holds. Even though \eqref{eq:dualLyapcontrol} is again bilinear, it becomes linear in terms of $(\rho, \rho \bu)$. Thus, the density function based method for control synthesis is a convex problem. 

\subsection{Sum of squares} \label{sec:SOS}
SOS optimization \cite{Topcu_10,Pablo_03,Pablo_01,Pablo_2000} is a relaxation of positive polynomial constraints appearing in polynomial optimization problems which are generally difficult to solve. SOS polynomials are in a set of polynomials which can be described as a finite linear combinations of monomials, i.e., $p = \sum_{i=1}^\ell d_i p_i^2$ where $p$ is a SOS polynomial; $p_i$ are monomials; and $d_i$ are coefficients. Hence, SOS is a sufficient condition for nonnegativity of a polynomial and thus SOS relaxation provides a lower bound on the minimization problems of polynomial optimizations. Using the SOS relaxation, any polynomial optmization problems with positive constraints can be formulated as SOS optimization as follows:
\begin{align} \label{eq:SOSOPT}
\begin{split}
    \min_{\bd} \, \mathbf{w}^\top \mathbf{d} \,\,\, \mathrm{s.t.} \,\,\, p_s(\bx,\bd) \in \Sigma[\bx], \, p_e(\bx;\bd) = 0,
\end{split}
\end{align}
where $\Sigma[\bx]$ denotes SOS set; $\mathbf{w}$ is weighting coefficients; $p_s$ and $p_e$ are polynomials with coefficients $\bd$. The problem in~\eqref{eq:SOSOPT} is translated into Semidefinite Programming (SDP)~\cite{Pablo_03, Laurent_2009}. There are readily available SOS optimization packages such as SOSTOOLS~\cite{sostools} and SOSOPT~\cite{Seiler_2013} to solve~\eqref{eq:SOSOPT}.

\subsection{Linear Koopman and Perron-Frobenius Operators}
For a dynamical system, $\dot \bx=\bF(\bx)$, there are two different ways of linearly lifting the finite dimensional nonlinear dynamics from state space to infinite dimension space of functions, ${\cal F}$, namely Koopman and Perron-Frobenius operators. Denote the solution of system~\eqref{eq:controlaffine} by $\phi_t(\bx)$. The definitions of these operators along with the infinitesimal generators of these operators are defined as follows.
\begin{definition}[Koopman Operator]  $\mathbb{K}_t :{\cal F}\to {\cal F}$ for dynamical system~\eqref{eq:controlaffine} is defined as 
\[[\mathbb{K}_t \varphi](\bx)=\varphi(\phi_t(\bx)),\;\;\varphi\in {\cal F}, \;\;\;t\geq 0. 
\]
The infinitesimal generator for the Koopman operator is
\begin{eqnarray}
\lim_{t\to 0}\frac{(\mathbb{K}_t-I)\varphi}{t}=\bF(\bx)\cdot \nabla \varphi(\bx)=:{\cal K}_{\bF} \varphi \label{K_generator}
\end{eqnarray}
\end{definition}

\begin{definition}[Perron-Frobenius Operator]  $\mathbb{P}_t:{\cal F}\to {\cal F}$ for dynamical system~\eqref{eq:controlaffine} is defined as 
\[[\mathbb{P}_t \psi](\bx)=\psi(\phi_{-t}(\bx))\left|\frac{\partial \phi_{-t}(\bx) }{\partial \bx}\right|,\;\;\psi\in {\cal F}, \;\;\;t\geq 0 \]
where $\left|\cdot \right|$ stands for the determinant. The infinitesimal generator for the P-F operator is given by 
\begin{eqnarray}\lim_{t\to 0}\frac{(\mathbb{P}_t-I)\psi}{t}=-\nabla \cdot (\bF(\bx) \psi(\bx))=: {\cal P}_{\bF}\psi \label{PF_generator}
\end{eqnarray}
\end{definition}
These two operators are dual to each other where the duality is expressed as follows.
\begin{eqnarray}
\int_{\mathbb{R}^n}[\mathbb{K}_t \varphi](\bx)\psi(\bx)d\bx=
\int_{\mathbb{R}^n}[\mathbb{P}_t \psi](\bx)\varphi(\bx)d\bx
\end{eqnarray}

\section{Data-driven control synthesis} \label{sec:main}
We are interested in data-driven control synthesis for multivariate nonlinear dynamics\footnote{we use bold symbols to denote column vectors unless it is specified as a row vector or a matrix.}:
\begin{equation} \label{eq:model}
    \dot{\mathbf{x}} = \bF(\bx) + \bG(\bx) \bu,
\end{equation}
where state $\bx\in \mR^n$ and control inputs $\bu$; and $\bF$ represents open-loop dynamics; and $\bG(\bx)=(\bG_1(\bx),\ldots,\bG_m(\bx))$  constitutes feedback control loop corresponding to control inputs $\bu = [u_1, \ldots, u_m]^\top$. The explicit description of $\bF$ and $\bG$ are not available, but we have access to a set of sample trajectories generated from this system \eqref{eq:model}. Our goal is a state feedback strategy $\bu$ that globally stabilizes~\eqref{eq:model}.

\subsection{Density function approach reformulation} \label{sec:density}
Based on the density function method, \cite{PraParRan04} proposed an implementable algorithm using SOS. In particular, the parameterization
	\begin{equation}\label{eq:density}
		\rho(\bx) = \frac{a(\bx)}{b(\bx)^\alpha}, \quad \rho(\bx) \bu(\bx) = \frac{\bc(\bx)}{b(\bx)^\alpha},
	\end{equation}
where $a$ and $\bc=[c_1,\ldots,c_m]^\top$ are polynomials, $b$ is a positive polynomial (positive at $\bx\neq 0$), and $\alpha$ is a sufficiently large number such that the integrability condition in Theorem \eqref{thm:dual} holds. 

With this parametrization \eqref{eq:density}, \eqref{eq:dualLyapcontrol} becomes 
	\begin{eqnarray*}
		&&\nabla \cdot (\rho (\bF+ \bG \bu)) = \nabla \cdot [\frac{1}{b}(\bF a+\bG \bc)]
		\\\!\!&=&\!\! \frac{1}{b^{\alpha+1}}[b\nabla\cdot(\bF a+\bG \bc)-\alpha \nabla b\cdot(\bF a + \bG \bc)]
		\\\!\!&=&\!\!  \frac{1}{b^{\alpha+1}}[(1+\alpha)b\nabla\cdot(\bF a+\bG \bc)-\alpha \nabla\cdot (b \bF a + b \bG \bc )]
	\end{eqnarray*}
The positive polynomial $b$ can be chosen as a quadratic control Lyapunov function for the linearized dynamics at the origin $\bx=0$ \cite{PraParRan04}. The control synthesis then becomes finding polynomials $a$ and $\bc$ such that 
\begin{equation} \label{eq:stability}
(1+\alpha) b \nabla \cdot (\mathbf{F}a + \mathbf{G} \bc) - \alpha \nabla \cdot ( b \mathbf{F} a + b \mathbf{G} \bc) > 0,
\end{equation}
which is clearly a standard SOS problem.

\subsection{Data-driven approximation of linear operators} \label{sec:data-driven}
The fundamental object of interest in the data-driven control synthesis is the approximation of the infinitesimal generator of P-F operator shown in~\eqref{PF_generator} corresponding to vector fields $\bF$ and $\bG$ affine in control system~\eqref{eq:model}. For the finite dimensional approximate representation of inequality~\eqref{eq:stability}, we will approximate the divergence terms, i.e., $\nabla \cdot (\bF \, \cdot \, )$ and $\nabla \cdot (\bG_i \, \cdot \, )$ for $i=1,\ldots, m$, using Koopman and P-F generators. We adopt the technique from \cite{Huang_2018,huang2020data} for the approximation of these two generators. In particular, data generated from the control system~\eqref{eq:model} with zero input and unit step inputs for each control input is used for the approximation of the generators  ${\cal P}_\bF$ and ${\cal P}_{\bF+\bG_i}$ respectively. Using linearity property, the infinitesimal generator for $\bG_i$ i.e., ${\cal P}_{\bG_i}$ is approximated from ${\cal P}_{\bF+\bG_i}-{\cal P}_{\bF}={\cal P}_{\bG_i}$. Using similar argument, it also follows that 
\begin{eqnarray}{\cal K}_{\bG_i}={\cal K}_{\bF+\bG_i}-{\cal K}_\bF,\;\;i=1,\ldots,m \label{linear_gen}
\end{eqnarray}
Furthermore, we notice that the P-F generator for vector field $\bF$ can be written as 
\begin{eqnarray}
-{\cal P}_\bF \psi \hspace{-0.02in} = \hspace{-0.02in} \nabla\cdot(\bF \psi) \hspace{-0.02in} = \hspace{-0.02in} \bF \cdot \nabla \psi+\nabla\cdot \bF \psi \hspace{-0.02in} = \hspace{-0.02in} {\cal K}_\bF \psi+\nabla \cdot \bF \psi\label{PF_gen_approx}
\end{eqnarray}
This allows us to approximate the P-F generator using algorithm known for the approximation of Koopman operator such as Extended Dynamics Mode Decomposition (EDMD). We show that the multiplication operator corresponding to $\nabla \cdot \bF$ in~\eqref{PF_gen_approx} can also be approximated using the approximate Koopman operator.
\begin{figure*}[h]
\centering
\includegraphics[scale=0.2]{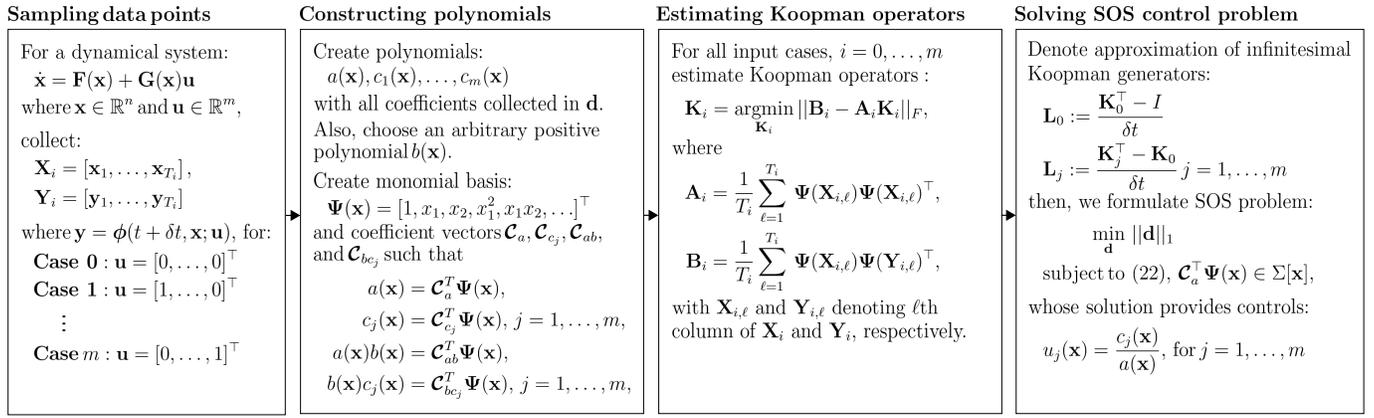}
\caption{Summary of the steps in our proposed algorithm described in Section~\ref{sec:main}.}
\label{fig:process}
\end{figure*}

For the data-driven approximation, let $\boldsymbol{\phi}(t,\mathbf{x; u})$ denote a solution of~\eqref{eq:model} at time $t$ starting from $\mathbf{x}$ with control input $\bu$. First, we collect time-series data from the dynamical system in~\eqref{eq:model} by injecting different control inputs: i) zero control inputs (i.e., $\bu=0$), and ii) unit step control inputs, i.e., $\bu=\mathbf{e}_j$ for $j=1,\ldots,m$ for a finite time horizon with sampling step $\delta t$, where $\be_j \in \mathbb{R}^m$ denotes unit vectors (i.e., $j$th entry of $\mathbf{e}_j$ is 1, otherwise 0). The time-series data of the system responses corresponding to each control input case are collected in:
\begin{align} \label{eq:data_matrix}
    \mathbf{X}_i = \left [ \mathbf{x}_1, \ldots, \mathbf{x}_{T_i} \right ], \,\, \mathbf{Y}_i = \left [ \mathbf{y}_1, \ldots, \mathbf{y}_{T_i} \right ],
\end{align}
with $i=0,1,\ldots,m$ for zero and unity control inputs, where $\mathbf{y}=\boldsymbol{\phi}(t+\delta t, \mathbf{x; u})$; and $T_i$ are the number of time-series data points collected for each input case. The samples in $\mathbf{X}_i$ do not have to be from a single trajectory; $\mathbf{X}_i$ can be a concatenation of multiple experiment/simulation trajectories.

We construct a polynomial basis denoted by 
\begin{align} \label{eq:basis functions}
\bPsi(\bx) = [\psi_1(\bx), \ldots, \psi_Q(\bx)]^\top
\end{align}
as a vector of monomials up to $q$th order. The total number of monomials in the basis, $Q={n+q \choose q}$. Using the EDMD algorithm, the Koopman operator, $\mathbb{K}_i$\footnote{For notational simplicity, we do not explicitly denote the Koopman operator dependence on the sampling time $\delta t$ i.e., $\mathbb{K}_{\delta t}$.} for $i=0,1,\ldots, m$ corresponding to zero input and step inputs ${\bf u}={\bf e}_j$ for $j=1,\ldots, m$ is approximated as \cite{WilKevRow15}:
\begin{align} \label{eq:finite Koopman}
    \mathbb {K}_i \approx \bK_i = \underset{\bK_i}{\mathrm{argmin}} \, ||\bB_i - \bA_i \bK_i||_F,
\end{align}
where $\bK_i$ are the estimated Koopman operator matrices for~$\mathcal{K}_i$,
\begin{align*}
        \mathbf{A}_i &= \frac{1}{T_i}\sum_{\ell=1}^{T_i} \, \boldsymbol{\Psi}(\bX_{i,\ell}) \boldsymbol{\Psi}(\bX_{i,\ell})^\top,\\
        \mathbf{B}_i &= \frac{1}{T_i}\sum_{\ell=1}^{T_i} \, \boldsymbol{\Psi}(\bX_{i,\ell}) \boldsymbol{\Psi}(\bY_{i,\ell})^\top,
\end{align*}
 and $\bX_{i,\ell}$ and $\bY_{i,\ell}$ denote $\ell$th column of $\bX_i$ and $\bY_i$, respectively. 
The solution of~\eqref{eq:finite Koopman} is explicitly known, $\mathbf{K}_i = \mathbf{A}_i^\dagger \mathbf{B}_i$, where $\dagger$ stands for pseudo-inverse.  The Koopman generator for vector field, $\bF$, can now be approximated as 
\begin{eqnarray} \label{eq:L0} 
{\cal K}_{\bF}\approx \frac{{\bf K_0}-I }{\delta t}=:{\bL}_0. \label{K_fapprox}
\end{eqnarray}
We approximate the multiplication operator corresponding to the divergence of vector field $\bf F$ as follows
\begin{eqnarray}
\nabla \cdot {\bf F} =\nabla \cdot [{\cal K}_0 x_1,\ldots, {\cal K}_0 x_n]^\top \approx \nabla \cdot(\bC_x^\top \bL_0 \bPsi) \label{approx_divergence}
\end{eqnarray}
where $\bC_x$ is a coefficient vector corresponding to the original states in the basis function $\bPsi$  i.e., $\bx = \bC_x^\top \bPsi$. Since, $\bPsi$ are assumed to be monomials basis, we can extract $\bx$ from $\bPsi$. Using linearity property of the generator in~\eqref{linear_gen}, we can approximate the Koopman generator corresponding to vector field $\bG_j$ for $j=1,\ldots, m$ as 
\begin{eqnarray}
{\cal K}_{\bG_j}\approx \frac{\bK_j-\bK_0}{\delta t}=:{\bL}_j,\;\;\;j=1,\ldots, m\label{K_gapprox}
\end{eqnarray}
Similarly following (\ref{approx_divergence}), the multiplication operator corresponding to the divergence of vector fields $\bG_j$ are approximated as 
\begin{eqnarray}\nabla\cdot (\bG_j)\approx \nabla \cdot(\bC_x^\top \bL_j \bPsi),\;\;j=1,\ldots,m.\label{divg_approx}
\end{eqnarray}
Finally, combining~\eqref{PF_gen_approx}, and~\eqref{eq:L0}--\eqref{divg_approx}, we have the following approximation for the infinitesimal generators for the P-F operators corresponding to the vector fields $\bF$ and $\bG_j$,~$\forall j$:
\begin{eqnarray} \label{eq:PF operators}
{\bP}_0=\bL_0+\nabla \cdot(\bC_x^\top \bL_0 \bPsi) {\bf I},\;\;\;{\bP}_j=\bL_j+\nabla \cdot(\bC_x^\top \bL_j \bPsi) {\bf I}
\end{eqnarray}

\subsection{Convex Control Synthesis: Combining SOS with Koopman} \label{sec:algorithm}
In this section, we formulate convex control synthesis using SOS optimization and Koopman operator described in previous sections.

First, we create polynomials $a(\bx)$, and $\bc(\bx) = [c_1(\bx),\ldots,c_m(\bx)]^T$ with degrees up to $q_a$, $q_{c_1},\ldots,q_{c_m}$, respectively. Coefficients of those polynomials are denoted by 
\begin{align*}
\bz_a = [\hat{a}_1, \ldots, \hat{a}_{Q_a}]^\top, \bz_{c_j} = [\hat{c}_{j,1} \ldots, \hat{c}_{j,Q_{c_j}}]^\top, j = 1, \ldots, m,
\end{align*}
where $Q_a={n+q_a \choose q_a}$ and $Q_{c_j} = {n+q_{c_j} \choose q_{c_j}}$. Subsequently, we manipulate $\bz_a$ and $\bz_{c_j}$ algebraically to create coefficient vectors $\bC_a$ and $\bC_{c_j}$ in terms of $\bPsi$ such that
\begin{align*}
    a(\bx)= \bC_a^\top \bPsi(\bx), \, c_j(\bx)=\bC_{c_j}^\top \bPsi(\bx), \, j = 1, \ldots, m.
\end{align*}
Similarly, let $\bC_{ab}$, $\bC_{bc_1}$, \ldots, $\bC_{bc_m}$ denote coefficient vectors of multiplications of polynomials, $a(\bx) b(\bx)$, $b(\bx) c_1(\bx)$, \ldots, $b(\bx) c_m(\bx)$, namely,
\begin{align*}
    a(\bx)b(\bx) = \bC_{ab}^\top \bPsi (\bx), b(\bx) c_j(\bx) = \bC_{bc_j}^\top \bPsi (\bx),\, j = 1, \ldots, m.
\end{align*}
$b(\mathbf{x})$ is an arbitrary positive polynomial appearing in~\eqref{eq:stability}. Note that the degree of the monomial basis $\bPsi(\bx)$ in~\eqref{eq:basis functions} should be larger than any other polynomials described above,
\begin{align*}
    \mathrm{deg}(\bPsi(\bx)) \geq \mathrm{max}(\mathrm{deg}(a(\bx) b(\bx)),\mathrm{deg}(b(\bx) c_j(\bx))),\forall j.
\end{align*}
Note that there is no systematic way to optimally choose the degree of polynomials, however we require higher order polynomials for $c(\bx)$ than $a(\bx)$ depending on the complexity of the underlying dynamics.
Now, using the approximation of the infinitesimal PF generators in~\eqref{eq:PF operators}, we restate the left-hand side of~\eqref{eq:stability} as below:
\begin{align} \label{eq:stability_approximation}
\begin{split}
&(1+\alpha) b(\bx) \left ( \bC_a^\top \bP_0 \bPsi(\bx) + \sum_{j=1}^m \bC_c^\top \bP_j \bPsi(\bx) \right )\\
&- b(\bx) \left( \bC_{ab} \bP_0 \bPsi(\bx) + \sum_{j=1}^m \bC_{bc_j}^\top \bP_j \bPsi(\bx) \right)
\end{split}
\end{align}
\begin{figure}[h]
\centering
\includegraphics[scale=0.20]{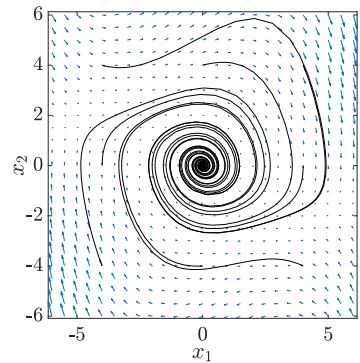}
\caption{Van der Pol oscillator stabilized by proposed method.}
\label{fig:VdP}
\end{figure}
The polynomial in~\eqref{eq:stability_approximation} is linear in terms of the coefficients of the polynomials, $a(\bx)$, $c_j(\bx)$, $j=1,\ldots,m$, hence we can solve SOS problem with \eqref{eq:stability_approximation} as a SOS constraint, given as~below:
\begin{align} \label{eq:SOSOPT2}
    \underset{\bd}{\mathrm{min}} \,\, ||\bd||_1 \,\, \mathrm{subject \, to} \,\, \eqref{eq:stability_approximation},\, \bC_a^\top \bPsi(\bx) \in \Sigma[\mathbf{x}],
\end{align}
where $\bd = [\bz_a^\top, \bz_c^\top]^\top$ and the objective function is $\ell_1$-norm minimization to promote sparsity and robustness of solution. The last term in~\eqref{eq:SOSOPT2} reflects the constraint, $\rho > 0$. Subsequent to solving~\eqref{eq:SOSOPT2}, we can construct a controller $u_j(\mathbf{x}) = c_j(\bx)/a(\bx)$, $j=1,\ldots,m$ to stabilize the dynamical system in~\eqref{eq:model}. The steps of the proposed method described here in Section~\ref{sec:main} is summarized in Fig.~\ref{fig:process}.

\begin{figure}[h]
\centering
\includegraphics[scale=0.21]{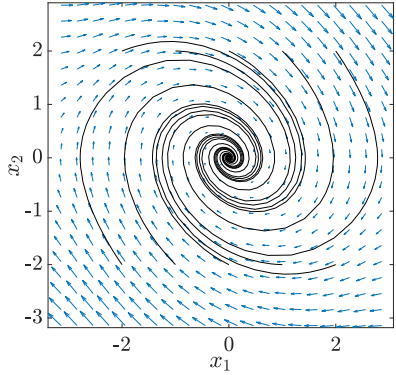}
\caption{Pendulum dynamics stabilized by proposed method.}
\label{fig:pendulum}
\end{figure}

\section{Numerical Case Studies} \label{sec:case study}
\subsection{Van der Pol Oscillator} \label{sec:VdP}
Dynamics of Van der Pol Oscillator is given as below~\cite{Ma_2019}:
\begin{align*}
    \dot{x}_1 = x_2, \,\,\, \dot{x}_2 = (1-x_1^2)x_2 - x_1 + u.
\end{align*}
We collect time-series data points of zero and unit step input responses in $\bX_{1 \sim 2}$ and $\bY_{1 \sim 2}$ shown in~\eqref{eq:data_matrix}, by doing repeated simulations. Simulation time spans from $0$ to $0.01$ [s] with time step $\delta t = 0.01$ [s], and we choose $10^4$ uniformly-distributed random initial points from $[x_1,x_2]=[-5,5]^2$. In this case, number of data points for each input response case, $T_1=9968$, $T_2=9970$. We choose $b(\bx) = \bx^\top \bx$, $a(\bx) = 1$, $\alpha = 6$, and also $c(\bx)$ to be a polynomial with degree from $1$ to $4$. Following the control synthesis described in Section~\ref{sec:main}, we have the solution, $c(\bx)= 0.9015 x_1^2 x_2 + 0.0251 x_2^3 + 0.0241 x_2^2 - 1.2505 x_2$. Following this, a synthesized control, $u(\bx) = c(\bx)$. Results of the control synthesis are shown in Fig.~\ref{fig:VdP} where trajectories starting from some initial points converge to the~origin.

\subsection{Non-Polynomial System Example: Inverted Pendulum}
Dynamics of a simple two-dimensional inverted pendulum is given as below:
\begin{align*}
    \dot{x}_1 = x_2, \,\,\, \dot{x}_2 = \mathrm{sin}x_1 - 0.5x_2 + u,
\end{align*}
which is non-polynomial due to a sinusoidal function. We collect time-series data points for zero and unit step inputs in $\bX_{1 \sim 2}$ and $\bY_{1 \sim 2}$, by doing repeated simulations, from $0$ to $0.001$ [s] with time step $\delta t= 0.001$ [s], starting from $10^4$ uniformly-distributed random initial points from $[x_1,x_2]=[-\pi,\pi]^2$. Number of data points for both input response cases, $T_1=T_2=10^4$. We choose $\alpha = 4$, $b(\bx) = \bx^\top \bx$, $a(\bx) = 1$, and $c(\bx)$ to be a polynomial with degree from $1$ to $3$. Following the proposed algorithm in Section~\ref{sec:main}, a control solution is computed, $u(\bx) = c(\bx) = 0.1553 x_1^3 - 1.9884 x_1$. Figure~\ref{fig:pendulum} shows trajectories of the dynamics with the synthesized control, starting from some initial points, demonstrating that the control solution from the proposed method can effectively stabilizes non-polynomial dynamical systems.

\begin{figure}[h]
\centering
\subfloat{\includegraphics[scale=0.17]{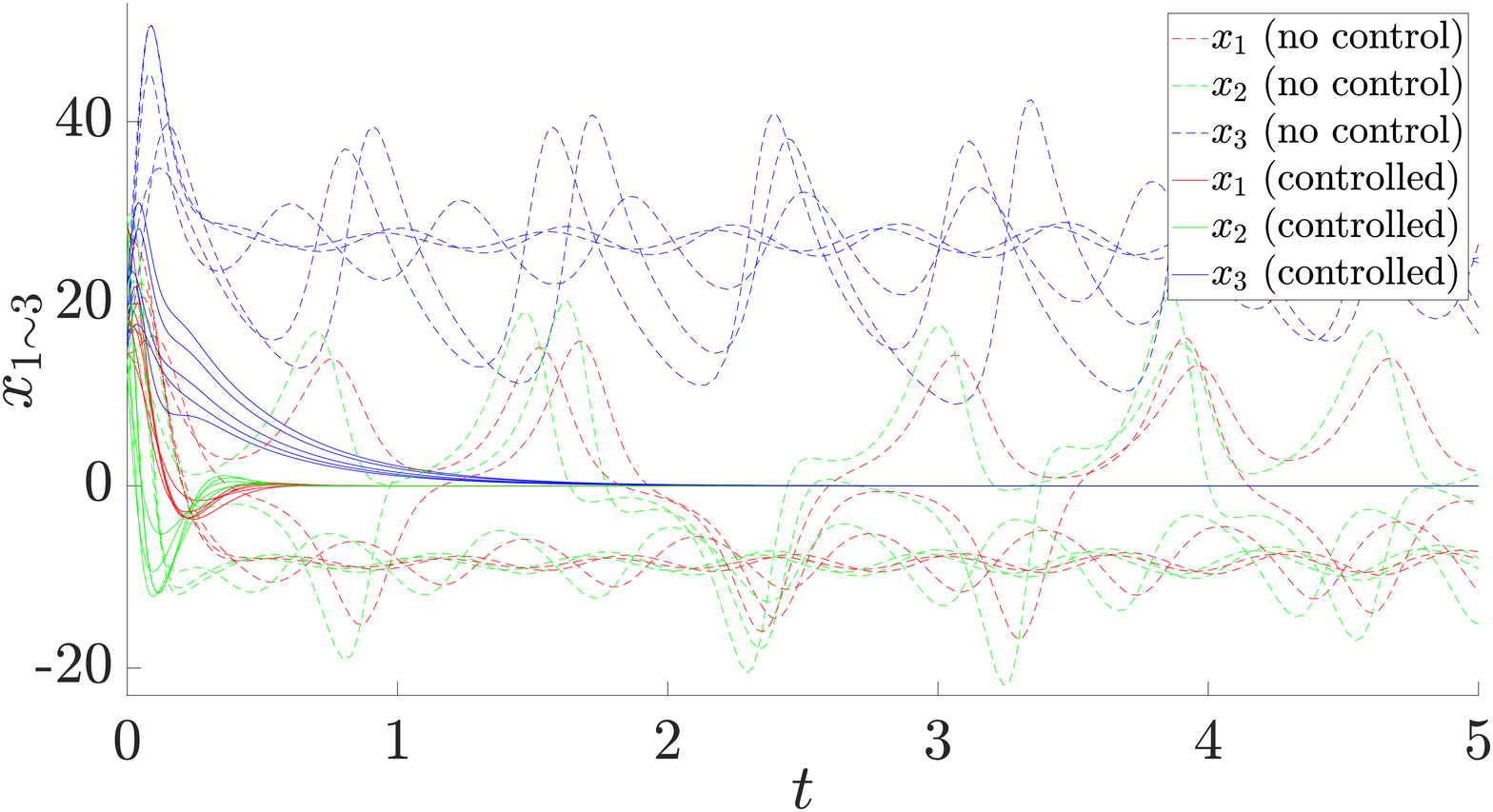}}\\ \vspace{-0.1in}
\subfloat{\includegraphics[scale=0.17]{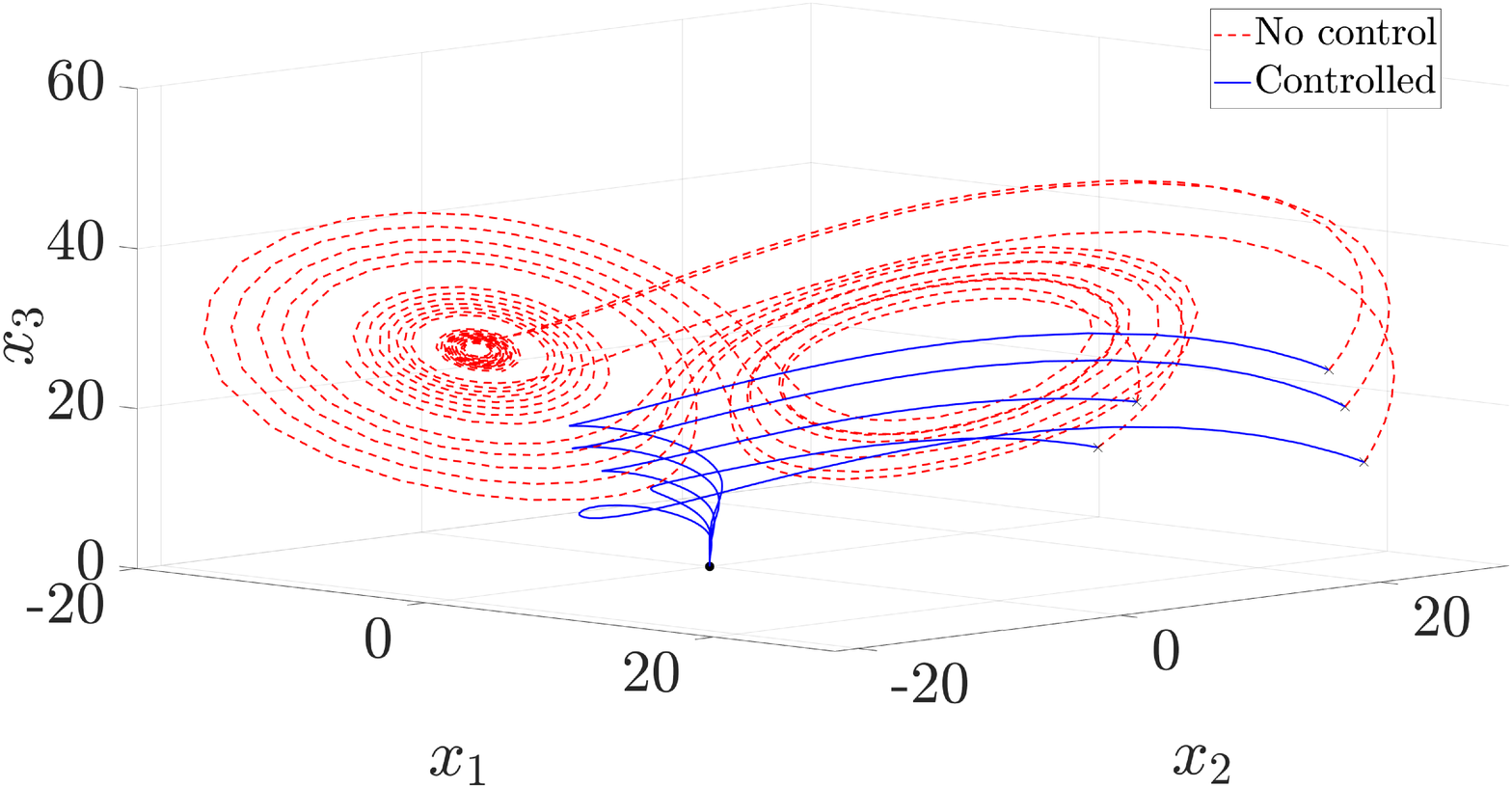}}
\caption{Lorenz attractor case result. Trajectories in states vs. time (top) and 3D plots (bottom) simulated from open-loop as well as controlled dynamics, starting from some disturbed initial points, converge to the origin while open-loop dynamics shows chaotic behavior.}
\label{fig:Lorenz}
\end{figure}
\vspace{-0.5cm}
\subsection{Lorenz System Dynamics}
Dynamics of Lorenz attractor is given by~\cite{Huang_2018}:
\begin{align*}
    \dot{x}_1 &= \sigma (x_2 - x_1),\\
    \dot{x}_2 &= x_1(\rho - x_3) - x_2 + u,\\
    \dot{x}_3 &= x_1x_2 - \beta x_3,
\end{align*}
where $\rho=28$, $\sigma = 10$, and $\beta = \frac{8}{3}$. We sample the time-series data points from repeated simulations, from $0$ to $0.001$ [s], with time step $\delta t = 0.001$ [s], and uniformly distributed initial points collected from $[x_1, x_2, x_3] = [-5 \times 5]^3$. The data points collected for all input cases, $T_1 = T_2 = 9945$. For the parameters of stability conditions, we choose $\alpha = 4$, $b(\bx) = \bx^\top \bx$, $a(\bx)=1$, and $c(\bx)$ to be a polynomial with degree from $1$ to $3$. Following the proposed method described in Section~\ref{sec:main}, we get the solution, $u(\bx) = c(\bx) = - 26.9591 x_1 - 6 x_2$, and the result of the control synthesis is depicted in Fig.~\ref{fig:Lorenz}, showing trajectories of the open-loop dynamics as well as the controlled dynamics, starting from different initial conditions. We can see that chaotic dynamics of the Lorenz attractor is stabilized to the origin by the control synthesized by our proposed method.

\subsection{Rigid Body Control}
In this case study, we investigate the dynamics of a rigid body system, which consists of six dynamical states and three control inputs~\cite{PraParRan04}:
\begin{align} \label{eq:rigid body}
\begin{split}
    \dot{\omega} &= \bJ^{-1} S(\bomega) \bJ \bomega + \bJ^{-1} \bu,\\
    \dot{\bpsi} &= \bH(\bpsi) \bomega,
\end{split}
\end{align}
where the angular velocity vector, $\bomega \in \mathbb{R}^3$; Rodrigues parameter vector, $\bpsi \in \mathbb{R}^3$; and control torque, $\bu \in \mathbb{R}^3$. We follow the same parameters $\bJ$, $\bS$, and $\bH$, as shown in~\cite{PraParRan04}.

\begin{figure}[h]
\centering
\subfloat{\includegraphics[scale=0.17]{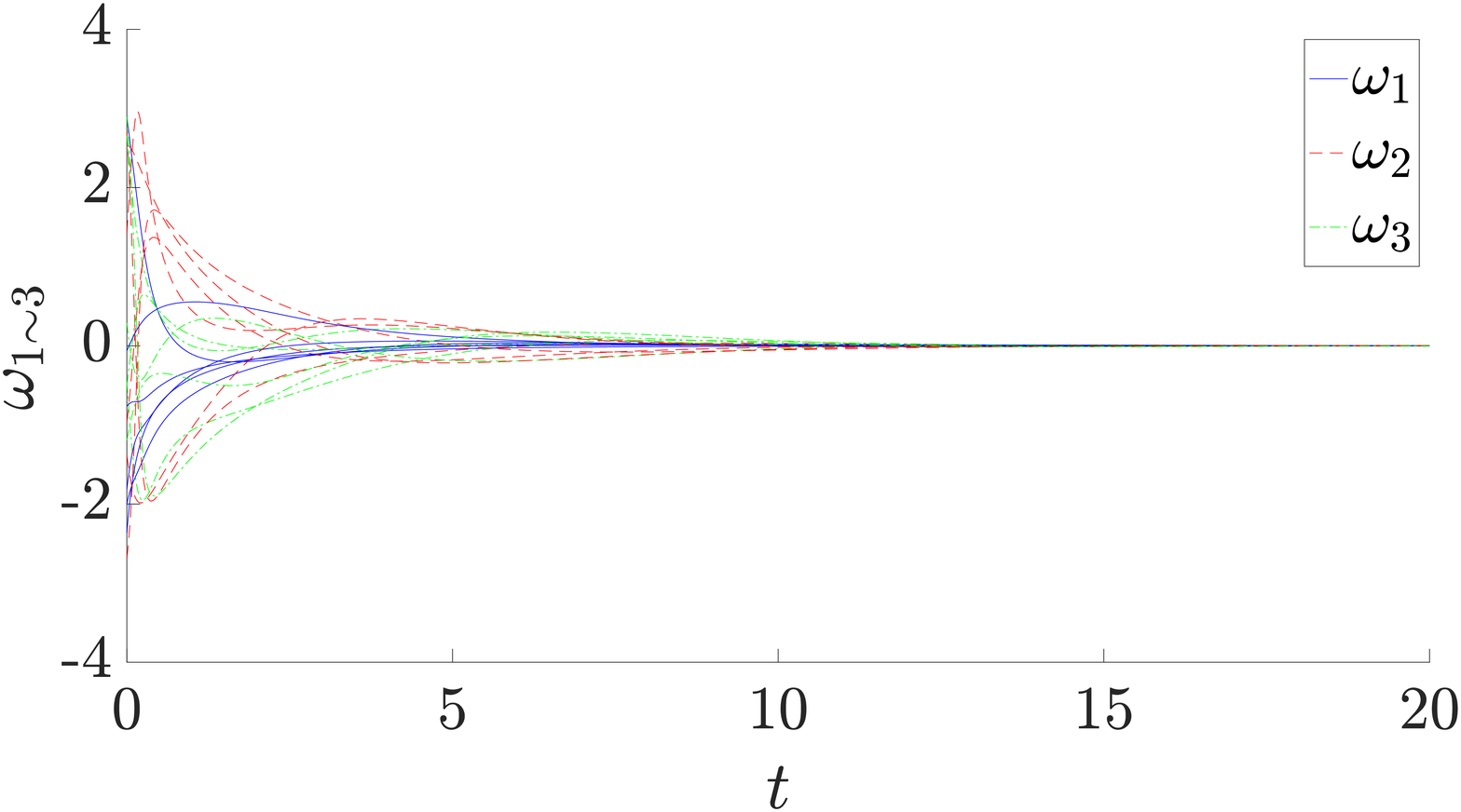}}\\ \vspace{-0.18in}
\subfloat{\includegraphics[scale=0.17]{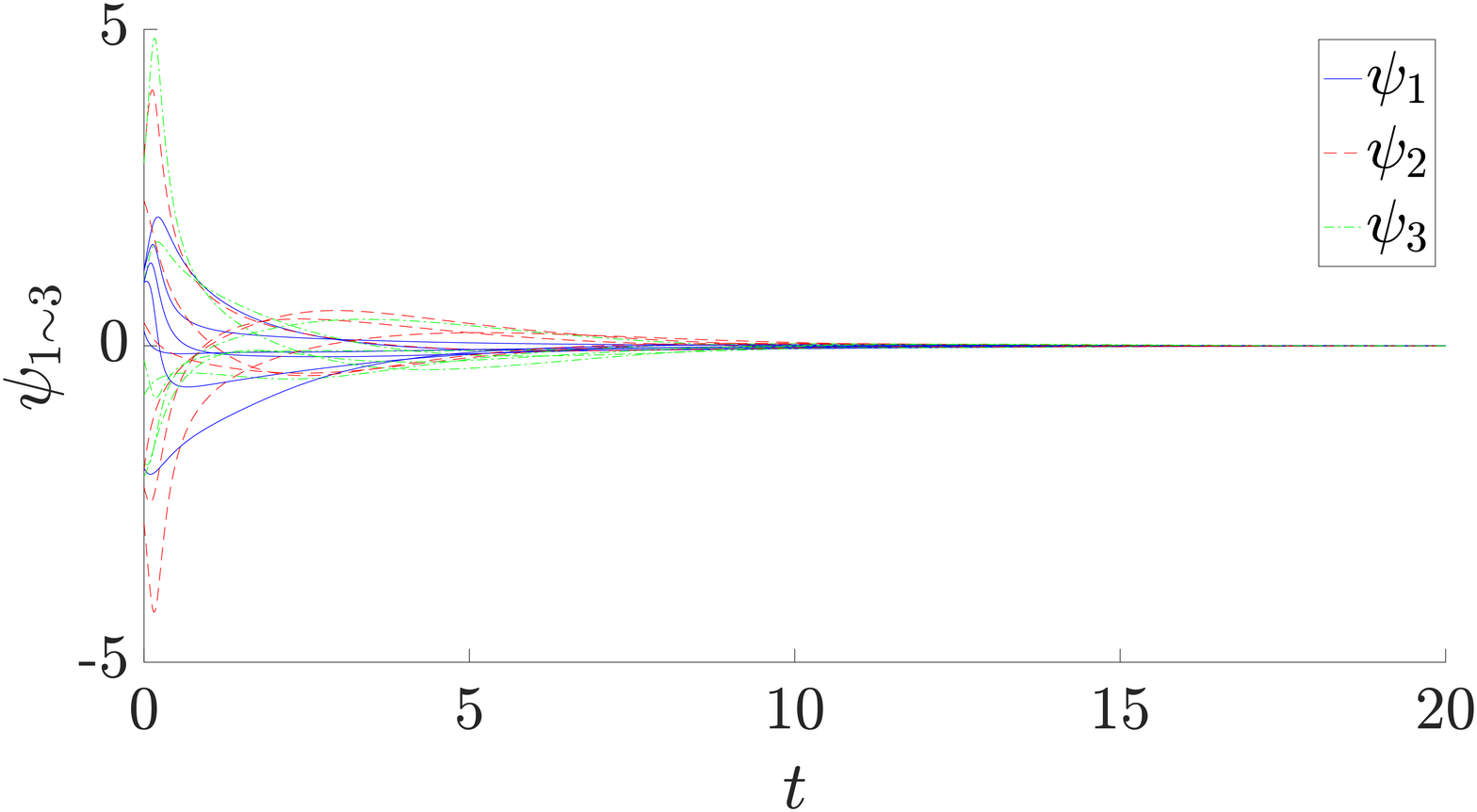}}
\caption{Result of stabilizing control synthesized from the proposed data-driven method for rigid body system.}
\label{fig:rigidbody}
\end{figure}

Time-series data points are sampled from repeated time-domain simulations for four control input cases, i.e., $\bu = 0$, $\bu = \be_{1 \sim 3}$. Simulation time spans from $0$ to $0.001$ [s] with time step $\delta t = 0.001$ [s], starting from uniformly distributed random initial points, $[\bomega^\top, \bpsi^\top] = [-3 \times 3]^6$. Each data matrix, $\bX_{1 \sim 4}$, $\bY_{1 \sim 4}$ has 9986 time-series data points. The parameters of stability formulation are chosen as $\alpha=4$, $a(\bx)=1$, and $c_j(\bx)$ to be a polynomial with degree from $1$ to $3$. Also, $b(\bx) = ||\bomega + \bpsi||^2 + ||\bpsi||^2$ which is known to be a CLF of the linearized dynamics of~\eqref{eq:rigid body} from~\cite{PraParRan04}. 
The resulting control $u_j=c_j(\bx)$, $j=1,...,3$. Figure~\ref{fig:rigidbody} shows trajectories of the states $\bomega_{1 \sim 3}$ and $\bpsi_{1 \sim 3}$, starting from some initial points, which demonstrates that the proposed method can deal with higher dimensional dynamical systems.

\section{Concluding Remark} \label{sec:conclusion}
A systematic convex optimization-based framework is provided for data-driven stabilization of control affine nonlinear systems. The proposed approach relies on a combination of SOS optimization methods and recent advances in the data-driven computation of the Koopman operator. Future research efforts will focus on data-driven optimal control of the nonlinear system and the robust counterpart of this work by exploiting the sample complexity of Koopman and P-F operator~\cite{CheVai18}.
\vspace{-0.1in}
\bibliographystyle{IEEEtran}
\bibliography{refs}

\begin{thebibliography}{10}
\providecommand{\url}[1]{#1}
\csname url@rmstyle\endcsname
\providecommand{\newblock}{\relax}
\providecommand{\bibinfo}[2]{#2}
\providecommand\BIBentrySTDinterwordspacing{\spaceskip=0pt\relax}
\providecommand\BIBentryALTinterwordstretchfactor{4}
\providecommand\BIBentryALTinterwordspacing{\spaceskip=\fontdimen2\font plus
\BIBentryALTinterwordstretchfactor\fontdimen3\font minus
  \fontdimen4\font\relax}
\providecommand\BIBforeignlanguage[2]{{%
\expandafter\ifx\csname l@#1\endcsname\relax
\typeout{** WARNING: IEEEtran.bst: No hyphenation pattern has been}%
\typeout{** loaded for the language `#1'. Using the pattern for}%
\typeout{** the default language instead.}%
\else
\language=\csname l@#1\endcsname
\fi
#2}}

\bibitem{Ran01}
A.~Rantzer, ``A dual to {L}yapunov's stability theorem,'' \emph{Systems \&
  Control Letters}, vol.~42, no.~3, pp. 161--168, 2001.

\bibitem{vaidya2008lyapunov}
U.~Vaidya and P.~G. Mehta, ``Lyapunov measure for almost everywhere
  stability,'' \emph{IEEE Transactions on Automatic Control}, vol.~53, no.~1,
  pp. 307--323, 2008.

\bibitem{rajaram2010stability}
R.~Rajaram, U.~Vaidya, M.~Fardad, and B.~Ganapathysubramanian, ``Stability in
  the almost everywhere sense: A linear transfer operator approach,''
  \emph{Journal of Mathematical analysis and applications}, vol. 368, no.~1,
  pp. 144--156, 2010.

\bibitem{das2018data}
A.~K. Das, B.~Huang, and U.~Vaidya, ``Data-driven optimal control using
  transfer operators,'' in \emph{IEEE Conference on Decision and Control
  (CDC)}.\hskip 1em plus 0.5em minus 0.4em\relax IEEE, 2018, pp. 3223--3228.

\bibitem{raghunathan2013optimal}
A.~Raghunathan and U.~Vaidya, ``Optimal stabilization using lyapunov
  measures,'' \emph{IEEE Transactions on Automatic Control}, vol.~59, no.~5,
  pp. 1316--1321, 2013.

\bibitem{WilKevRow15}
\BIBentryALTinterwordspacing
M.~O. Williams, I.~G. Kevrekidis, and C.~W. Rowley, ``A data-driven
  approximation of the {K}oopman operator: Extending dynamic mode
  decomposition,'' \emph{Journal of Nonlinear Science}, vol.~25, no.~6, pp.
  1307--1346, Jun 2015. [Online]. Available:
  \url{http://dx.doi.org/10.1007/s00332-015-9258-5}
\BIBentrySTDinterwordspacing

\bibitem{Mezic_2013}
\BIBentryALTinterwordspacing
I.~Mezić, ``Analysis of fluid flows via spectral properties of the koopman
  operator,'' \emph{Annual Review of Fluid Mechanics}, vol.~45, no.~1, pp.
  357--378, 2013. [Online]. Available:
  \url{https://doi.org/10.1146/annurev-fluid-011212-140652}
\BIBentrySTDinterwordspacing

\bibitem{Susuki_2016}
\BIBentryALTinterwordspacing
Y.~Susuki, I.~Mezic, F.~Raak, and T.~Hikihara, ``Applied koopman operator
  theory for power systems technology,'' \emph{Nonlinear Theory and Its
  Applications, IEICE}, vol.~7, no.~4, p. 430–459, 2016. [Online]. Available:
  \url{http://dx.doi.org/10.1587/nolta.7.430}
\BIBentrySTDinterwordspacing

\bibitem{Vaidya_2019}
P.~{Sharma}, B.~{Huang}, V.~{Ajjarapu}, and U.~{Vaidya}, ``Data-driven
  identification and prediction of power system dynamics using linear
  operators,'' in \emph{2019 IEEE Power Energy Society General Meeting
  (PESGM)}, 2019, pp. 1--5.

\bibitem{mauroy2016global}
A.~Mauroy and I.~Mezi{\'c}, ``Global stability analysis using the
  eigenfunctions of the {K}oopman operator,'' \emph{IEEE Transactions on
  Automatic Control}, vol.~61, no.~11, pp. 3356--3369, 2016.

\bibitem{korda2018linear}
M.~Korda and I.~Mezi{\'c}, ``Linear predictors for nonlinear dynamical systems:
  Koopman operator meets model predictive control,'' \emph{Automatica},
  vol.~93, pp. 149--160, 2018.

\bibitem{huang2020data}
B.~Huang, X.~Ma, and U.~Vaidya, ``Data-driven nonlinear stabilization using
  koopman operator,'' in \emph{The Koopman Operator in Systems and
  Control}.\hskip 1em plus 0.5em minus 0.4em\relax Springer, 2020, pp.
  313--334.

\bibitem{kaiser2017data}
E.~Kaiser, J.~N. Kutz, and S.~L. Brunton, ``Data-driven discovery of {K}oopman
  eigenfunctions for control,'' \emph{arXiv preprint arXiv:1707.01146}, 2017.

\bibitem{kaiser2019datadriven}
E.~Kaiser, J.~N. Kutz, and S.~L. Brunton, ``Data-driven approximations of
  dynamical systems operators for control,'' 2019.

\bibitem{Topcu_10}
U.~{Topcu}, A.~{Packard}, P.~{Seiler}, and G.~{Balas}, ``Help on sos [ask the
  experts],'' \emph{IEEE Control Systems Magazine}, vol.~30, no.~4, pp. 18--23,
  2010.

\bibitem{Pablo_03}
P.~A. Parrilo, ``Semidefinite programming relaxations for semialgebraic
  problems,'' \emph{Mathematical Programming}, vol.~96, pp. 293--320, May 2003.

\bibitem{Pablo_01}
P.~A. Parrilo and B.~Sturmfels, ``Minimizing polynomial functions,'' 2001.

\bibitem{Pablo_2000}
P.~A. Parrilo, ``Structured semidefinite programs and semialgebraic geometry
  methods in robustness and optimization,'' Ph.D. dissertation, California
  Institute of Technology, May 2000.

\bibitem{Laurent_2009}
M.~Laurent, \emph{Sums of Squares, Moment Matrices and Optimization Over
  Polynomials}.\hskip 1em plus 0.5em minus 0.4em\relax New York, NY: Springer
  New York, 2009, pp. 157--270.

\bibitem{sostools}
A.~{Papachristodoulou}, J.~{Anderson}, G.~{Valmorbida}, S.~{Prajna},
  P.~{Seiler}, and P.~A. {Parrilo}, \emph{{SOSTOOLS}: Sum of squares
  optimization toolbox for {MATLAB}}, \texttt{http://arxiv.org/abs/1310.4716},
  2013.

\bibitem{Seiler_2013}
P.~Seiler, ``Sosopt: A toolbox for polynomial optimization,'' 2013.

\bibitem{PraParRan04}
S.~Prajna, P.~A. Parrilo, and A.~Rantzer, ``Nonlinear control synthesis by
  convex optimization,'' \emph{IEEE Transactions on Automatic Control},
  vol.~49, no.~2, pp. 310--314, 2004.

\bibitem{Huang_2018}
B.~{Huang}, X.~{Ma}, and U.~{Vaidya}, ``Feedback stabilization using koopman
  operator,'' in \emph{2018 IEEE Conference on Decision and Control (CDC)},
  2018, pp. 6434--6439.

\bibitem{Ma_2019}
X.~{Ma}, B.~{Huang}, and U.~{Vaidya}, ``Optimal quadratic regulation of
  nonlinear system using koopman operator,'' in \emph{2019 American Control
  Conference (ACC)}, July 2019, pp. 4911--4916.

\bibitem{CheVai18}
Y.~Chen and U.~Vaidya, ``Sample complexity for nonlinear stochastic dynamics,''
  in \emph{2019 American Control Conference (ACC)}.\hskip 1em plus 0.5em minus
  0.4em\relax IEEE, 2019, pp. 3526--3531.

\end{thebibliography}

\end{document}